**Possible scenarios that the ¨New Horizons¨ spacecraft may find in its close encounter with Pluto**

Héctor Javier Durand-Manterola and Héctor Pérez-de-Tejada
Instituto de Geofísica, Universidad Nacional Autónoma de México (UNAM)
hdurand_manterola@yahoo.com

Abstract

Next year (2015) the ¨New Horizons¨ spacecraft will have **a** close encounter with Pluto. In the present study we discuss some possibilities regarding what the spacecraft may encounter during its approach to Pluto. Among them we should include: the presence of geological activity due to heat generated by tides; the unlikely presence of an intrinsic magnetic field; the possibility of a plasmasphere and a plasmapause; the position of an ionopause; the existence of an ionospheric trans-terminator flow similar to that at Venus and Mars; and the presence of a Magnus force that produces a deflection of Pluto's plasma wake. This deflection oscillates up and down in its orbit around the sun.

1 Introduction

In 1976 frozen $CH_4$ was identified in the surface of Pluto (Cruikshank et al., 1976; Hansen y Paige, 1996). It was known that Pluto has a bright surface, and since methane ice tends to darken when subjected to **r**adiation, it was proposed that sublimation and transport of methane keep fresh the surface (Stern et al., 1988), thus implying the presence of an atmosphere. Through stellar occultation a tenuous atmosphere was detected in 1988, and assuming that nitrogen was its main constituent, a pressure between 0.2 and 0.5 Pa was calculated (Elliot et al., 1989; Hubbard et al., 1990; Elliot y Young, 1991; Hansen y Paige, 1996). Finally**,** in 1992 the spectral signatures of nitrogen and carbon monoxide were identified, the two species occurring in solid state (Owen et al., 1992; Hansen y Paige, 1996).

Comparisons have been made between Triton and Pluto since both bodies have similar density, radius, rotation speed, surface gravity, atmospheric pressure and albedo. The volatile inventory of both, include $N_2$, $CH_4$ y CO. Both bodies have a thin atmosphere whose main component is nitrogen. Both have bright poles (Hansen y Paige, 1996).
The nature of the interaction of the solar wind with the ionosphere of Pluto depends on the amount of material the atmosphere lost. If Pluto has a large atmospheric escape flux interaction will be comet type. On the other hand if it has a weak atmospheric escape flux interaction will be Venus type (Bagenal and McNutt, 1989)

The New Horizons spacecraft will reach Pluto in July 2015 and will be the first approach to a dwarf planet in the Kuiper belt and the second visit to a dwarf planet after the visit of the ¨Dawn spacecraft¨ to Ceres in February of that year. In this study, we explore some possibilities about what the ¨New Horizons may find in its approach to Pluto.

2 Scenarios

2.1 Intrinsic Magnetic field in Pluto

If Pluto has a fluid inner layer and this layer is conductive, then Pluto may have a dynamo that generates a magnetic field. Durand-Manterola (2009) obtained the following empirical equation, which relates the mass of the planet (Mp), the conductivity of the conductive layer (σ) and the period of rotation of the body ($P_e$), with the planetary magnetic moment (M):

$$M = 1\times10^{-5}\left[\frac{M_p \sigma}{P_e}\right]^{1.106} \quad (2.1.1)$$

Using this equation we can calculate the magnetic field present at Pluto. If we assume that Pluto has a (very unlikely) molten iron core, then with the appropriate values (Table 1), the equation (2.1.1) gives the following value of the magnetic moment: $5.34\times10^{18}$ Amp-m$^2$, which is one order of magnitude smaller than the magnetic moment of Mercury.

Pluto is like the icy moons of the giant planets. Of these, Ganymede is the only with an intrinsic magnetic field. It has been suggested that Ganymede's magnetic field is generated in an underground saltwater ocean (Kivelson et al., 2002). If this is also the case for Pluto its conductivity will be less than that assuming iron. With the maximum value of seawater conductivity (Table 1) equation (2.1.1) gives a magnetic moment of $1.03\times10^{14}$ Amp-m$^2$.

If Pluto has an intrinsic magnetic field, then it has a magnetopause, which in the subsolar point, will be at the distance (Eq. (12) in Durand-Manterola, 2009, corrected):

$$R_{ss} = \left[\frac{\mu_0}{32\pi^2}\frac{M^2}{\rho V^2}\right]^{1/6} \quad (2.1.2)$$

where M is the dipolar magnetic moment, $\mu_0$ is the permeability of free space and $\rho$ and V are the density and velocity of the solar wind.

Applying the appropriate values (Table 1) in this equation and taking into account an iron core (which could be very unlikely), the subsolar point of the magnetopause will be 5768 km from the centre of Pluto, i.e., 4.95 Plutonian radii.

If the magnetic field is generated by an ocean of saltwater, equation (2.1.2) leads to $R_{SS}$ ~161.7 km which is an impossible value since it is smaller than Pluto's radius. In this case the magnetopause would be by the surface of Pluto or at the ionopause. Under such circumstances the presence of a magnetic field could only be verified in the wake where the magnetospheric plasma is not compressed.

About 50 minutes after its closest approach to Pluto the ¨New Horizons¨ will move through its wake. However, the spacecraft does not have a magnetometer and thus a magnetopause only will be inferred from observations of a discontinuity in the plasma properties such as a jump in density. On Earth the density and the temperature of the magnetospheric plasma gradually decrease within the magnetotail, being the highest at the magnetopause (Rosenbauer et al., 1975). If Pluto's magnetosphere is similar to that at Earth, we should expect such variations in density and temperature.

Before reaching the magnetopause, the solar wind will be influenced by a shock wave, and thus equation (2) will not be applicable since it involves a balance between the internal magnetic force and the kinetic energy of the free solar wind. However, when the shock wave slows down the solar wind it is compressed together with the convected magnetic field thus increasing the magnetic pressure (the kinetic pressure is converted into magnetic pressure). Therefore, a pressure balance is set between the magnetic pressure of the solar wind and the magnetic pressure of the planet. As long as the magnetic pressure of the solar wind is equal to the kinetic pressure, then (2.1.2) is applicable.

2.1.1 Plasmasphere and plasmapause

Inside the magnetospheres of planets there is a plasma zone corotating with the planet, the plasmasphere, in which the plasma provided by the ionosphere is driven by the corotation electric field. The distance to the boundary of this region, or plasmapause can be calculated by equating the centripetal force to the sum of the gravitational and the convective forces. From this we obtain:

$$r = \left[\frac{q\mu_0 M}{4\pi \omega m} + \frac{GM_P}{\omega^2}\right]^{1/3} \quad (2.1.3)$$

Assuming that there are only protons in the Plutonian plasmasphere and using the values of Table 1 we find that the corotation zone has a radius of 19,100 km, for the lower magnetic moment; and 238,000 km for the larger magnetic moment. In either case the radius of the corotation zone is greater than the distance to the subsolar magnetopause. Therefore, the whole frontal magnetosphere is corotating with Pluto.

For Pluto has a magnetic field requires that inside there is a conductive liquid layer. For this to happen requires an internal heat source. One possible source would tidal heating.

2.2 Tidal heating of the planetary body

Pluto and its moon Charon have trapped rotation, i.e., both bodies show each other the same hemisphere. On the other hand the orbit of Charon is circular (and therefore so is Pluto's orbit around their common centre of mass) (Buie et al., 2012). For these two conditions it is not expected that there is an exchange of energy, via tides, between the two bodies.

However, solar gravity force modifies this view since Charon may be in conjunction, in opposition, square and all intermediate positions with respect the Sun, thus both bodies will exert on Pluto a continuous changing tidal force that implies a continuous, and periodic, deformation of Pluto's body (and of course Charon's body). This continuous exchange of energy in tidal forces produces heat due to dissipation of tidal energy. The heat flow may manifest itself in cryo-volcanic activity on the surfaces of both bodies, or in the presence of a molten layer in their interior, or both. This happens only in certain positions in the orbit of Pluto around the Sun (see Figure 1). As the axis of rotation of the planet is lying almost in the plane of its orbit then the plane of the orbit of Charon is almost perpendicular to the orbital plane. As this plane is fixed respect to the background stars this means that there will be times when the plane of the orbit of Charon will be perpendicular to the direction Sun-Pluto and at other times will be parallel. In the latter situation would be where would have tidal effects.

If $F_C$ is the gravitational force exerted on Pluto by Charon, and $F_S$ **is** the gravitational force exerted by the Sun on Pluto, then the ratio of $F_C$ to $F_S$ is:

$$\frac{F_C}{F_S} = \frac{M_C}{M_S} \frac{R_S^2}{R_C^2} \qquad (2.2.1)$$

where M are the masses and R the distances of the Sun and Charon to Pluto and the subscripts S and C correspond, respectively, to both bodies. Putting the appropriate values (Table 1) in equation (2.2.1) we obtain that the force exerted by Charon on Pluto is 39.3 times the force exerted by the Sun in the perihelion and 108.4 times in the aphelion.

At perihelion, when the Sun and Charon are in conjunction as seen from Pluto the force exerted will be the sum of the two forces, i.e. 40.3 times the force exerted by the Sun. Three days later when Charon is in opposition to the Sun, the force exerted on Pluto is the difference between both forces, i.e., 38.3 times the force applied by the Sun. The difference between both positions is ~5 %. At aphelion, during the conjunction, the sum of the Sun and Charon forces is 109.4 times the force of the Sun, and in opposition 107.4 times **(T**his difference is 1.8 %**)**. The heat released by the deformation produced by these differential forces may aid in the sublimation of $N_2$ and $CH_4$, which are solid at the surface, and thus can help the formation of the atmosphere. Moreover, since Charon is less massive than Pluto, such dissipation may explain the fact that the surface of Charon is warmer than Pluto (Marcialis et al., 1987).

2.3 Ionopause on Pluto

The temperature on Pluto's surface is 40 ± 2 °K (Strobel et al., 1996). Given that the atmosphere is formed mainly of $N_2$, then the thermal velocity is 150 m/s. According to the model of Strobel et al., (1996) the atmospheric temperature at 80 km altitude reaches between 100 y 130 °K. Taking these values, the thermal velocities are 240 m/s and 280 m/s, respectively. On the other hand, since the escape velocity on Pluto is 1220 m/s (Tholen et al., 2000) the thermal velocities have smaller values and thus Pluto's atmosphere remains trapped.

Ions were detected in the Plutonian atmosphere (Young et al., 2008), that is, Pluto has an ionosphere. For all the above, and in the case that Pluto does not have a magnetic field or that it has a weak magnetic field generated by an ocean of salt water, then there will be a ionopause. We can calculate the position of the ionopause with a pressure balance between the thermodynamic pressure of the ionospheric plasma and the ram pressure of the solar wind.

$$\rho V^2 \cos^2 \xi = n_i k T \qquad (2.3.1)$$

Where $\rho$ is the density of the solar wind, V is the solar wind speed in free space at the distance of Pluto, $\xi$ is the subsolar angle, $n_i$ and T are the particle density and the ionospheric plasma temperature, and k is the Boltzmann constant.

If the solar wind arrives perpendicular to the ionosphere, as in the subsolar point ($\xi = 0$), then all its momentum is used to compress the ionosphere and ionopause be formed where the pressure and the thermal pressure of the ionosphere are balanced. But if the wind gets tilted to the vertical at some point in the ionosphere ($\xi \neq 0$) then its momentum will have two components: a vertical which will compress the ionosphere, and an horizontal which will have no effect on it. In this case the wind ram pressure will be lower, then the pressure of the ionosphere is balanced at higher altitudes where the density is less.

Before reaching the ionosphere, the solar wind decelerates as it moves across the shock wave; then strictly the pressure balance will occur between the magnetic field pressure plus the ram pressure of the shocked solar wind and the thermal pressure of the ionosphere. Since the magnetic pressure and the shocked ram pressure are equivalent to the free stream solar wind kinetic pressure then equation (2.3.1) is applicable.

From equation (2.3.1) and solving for $n_i$, we have:

$$n_i = \frac{n_{VS} m_P V^2}{kT} \cos^2 \xi \qquad (2.3.2)$$

Thus leading to the particle density of the ionosphere at the ionopause.
Moreover, from the model of Ip et al. (2000) we can obtain the altitude corresponding to the particle density value; namely:

$$h = 2.2611 \times 10^3 \exp(-3 \times 10^{-4} n_i) \qquad (2.3.3)$$

Combining equations (2.3.2) and (2.3.3) lead to the ionopause altitude. Figure (2.3.1) shows values of this altitude as a function of solar zenith angle. The altitude at the subsolar point is 2147 km and 2261 km by the terminator, or measured from the planet centre, 3312 km at the subsolar point, and 3426 km by the terminator.

When the Sun-Pluto line is in the plane of Charon's orbit the solar wind may not reach the ionopause by the subsolar point thus allowing the Plutonian ionosphere to expand.

2.4 Trans-terminator flow

When there is a magneto-hydrodynamic interaction between the solar wind flow and Pluto's ionosphere (basically wave-particle interactions), there should be momentum transfer from the solar wind to the upper ionosphere, which generates a flow of ions from dayside to night side (Perez-de-Tejada, 1986). This type of flow was first discovered at Venus and is called the ionospheric trans-terminator flow (Knudsen et al., 1980). A similar trans-terminator flow is expected to occur at Pluto. We can calculate its speed from an equation that relates the conservation of momentum flux between the solar wind and the ionosphere, i.e.:

$$\pi r_i^2 \rho_{SW} V_{SW}^2 = \pi (r_i^2 - r_0^2) \rho_i V_i^2 \qquad (2.4.1)$$

Here $\rho_{SW} V_{SW}^2$ is the momentum flux density of the solar wind and $\rho_i V_i^2$ is the momentum flux density in the ionosphere, $r_i$ is the radius of the ionopause at the terminator and $r_0$ is the radius of Pluto. Solving for $V_i$ we have:

$$V_i = V_{SW} \left[ \frac{r_i^2}{r_i^2 - r_0^2} \frac{\rho_{SW}}{\rho_i} \right]^{1/2} \qquad (2.4.2)$$

Using the values of Table 1 we obtain 2.8 km/s, which is larger than the escape velocity of 1.22 km/s and thus the momentum transfer from the solar wind to the Pluto ionosphere should produce a significant loss of ionospheric material.

2.5 Magnus force at Pluto

When a sphere moves through a fluid and at the same time rotates, a dynamic force, is produced and is oriented away from the directional flow (Magnus, 1853). Since Pluto is also immersed in the solar wind and is a rotating sphere it could also produce a Magnus force. However, since the solar wind is non-collisional plasma, it could interact directly with Pluto's solid body and thus the Magnus force, which is a hydrodynamic effect, would not occur. However, since Pluto has an ionosphere or a plasmasphere, its interaction with the solar wind could produce a magneto-hydrodynamic interaction resulting in a Magnus force. If Pluto has an intrinsic magnetic field the plasma of all the magnetosphere co-rotates with the planet, as we see in section 2.1.1. If it does not have a magnetic field then the lower ionosphere co-rotates with the solid body and the high region flows from dayside to night side (trans-terminator flow). In this case the Magnus effect occurs between the trans-terminator flow and the internal rotating ionosphere, but the effect of deviation of the wake, is manifests even in the solar wind.

Previous work has suggested that the Magnus force has an effect on the plasma wake of Venus and Mars (Perez-de-Tejada, 2006; Perez-de-Tejada et al., 2009). Such force was used to account for the deviated direction of the Venus plasma wake as a result of the solar wind interaction with the Venus ionosphere (Pérez-de-Tejada, 2006, 2008). It is possible that similar conditions are also applicable to Pluto's wake.

Since the plane of the equator of Pluto has 119.61 ° inclined to the plane of the orbit (Tholen et al., 2000) then as it moves in its orbit around the sun there will be times when the rotation axis is perpendicular to the solar wind flow. At this time is when will be Magnus effect. The Magnus effect will produce a deviation of the wake of the planet relative to the direction Sun-Pluto. In other parts of the orbit, when the rotation axis is parallel to the solar wind, there is no Magnus effect and the wake will be aligned with the Sun-planet line. In some parts of the Pluto's orbit the Magnus force points to the solar north and in others it points to the solar south (Pérez-de-Tejada et al., 2014). Such change will produce a cyclic variation of the wake deviation along Pluto's orbit. At times it will be diverted to the north of the orbital plane and in others to the south.

The Magnus force applied to a rotating planet is:

$$f_M = 4\pi\rho V \omega R^3 \qquad (2.5.1)$$

Where ρ and V are the density and speed of the solar wind, and ω is the angular velocity of rotation of the planet. R is the radius of the corotating plasma (ionopause or magnetopause) but not the radius of the planet since the Magnus force arises from the magneto-hydrodynamic interaction between the solar wind and a planetary plasma (ionosphere or plasmasphere).

2.6 Angle of the wake

The geometry of the flow generated by the Magnus force causes a deflection angle of the plasma wake. If the rotation speed of Pluto's ionosphere (or plasmasphere) is $V_P$ and the velocity of the flow is $V_f$ then in the upper ionosphere (or in the magnetopause) where both velocities are in the same direction the total speed is:

$$V_1 = V_f + V_p \qquad (2.6.1)$$

And in the side where the two velocities are anti-parallel we have:

$$V_2 = V_f - V_p \qquad (2.6.2)$$

Consider two parcels of flow together in the sub-solar point. One of the parcels moves to the side where the rotation motion and the flow are parallel and the other moves where the flow and the rotation motion are anti-parallel. The positions with time of the two parcels around Pluto are:

$$x_1 = V_1 t_1 - \pi R = (V_f + V_p)t_1 - \pi R \qquad (2.6.3)$$

$$x_2 = V_2 t_2 + \pi R = -(V_f - V_p)t_2 + \pi R \qquad (2.6.4)$$

Where R is the radius of the corotating sphere (ionosphere or plasmasphere), and $t_1$ and $t_2$ are the travel times of parcels 1 and 2 in their trajectories. We take the coordinate x measured on the surface of the co-rotating sphere, and zero in the anti-solar point. Therefore, parcel 1 begins at $x_1$ = -πR and the x-coordinate increases. Meanwhile, parcel 2 starts at $x_2$ = πR and the x-coordinate decreases.

When the parcels encounter again, after moving around Pluto, their travel time is: $t_1 = t_2 = t$ and their position is: $x_1 = x_2 = x$. Then we put these values into equation (2.6.4) and solving for t we obtain:

$$t = \frac{x - \pi R}{V_p - V_f} \qquad (2.6.5)$$

Substituting this value in equation (2.6.3), and after some algebraic manipulation, we have:

$$x = \pi R \frac{V_p}{V_f} \qquad (2.6.6)$$

This is the distance where the two parcels encounter, measured from the anti-solar point. At this point the parcels separate from Pluto and form the plasma wake. If we divide the equation (2.6.6) by R we have the angle of the wake (in radians) from the Sun-Pluto direction:

$$\theta = \pi \frac{V_p}{V_f} \qquad (2.6.7)$$

Multiplying by 180/π we have this angle in degrees:

$$\theta = 180 \frac{V_p}{V_f} \qquad (2.6.7)$$

Due to the velocity of Pluto in its orbit around the sun the solar wind does not reach exactly on the Sun-Planet direction, but at an angle to the right (as seen from the Sun):

$$\varphi = arctg\left(\frac{V_o}{V}\right) \qquad (2.6.8)$$

Where $V_o$ is Pluto's orbital velocity and V is the free stream solar wind velocity. Due to the fact that the planes in which the angles θ and φ occur are almost perpendicular to each other they cannot be added together. While φ keeps nearly a steady value (φ = 0.6° for Pluto's mean orbital velocity) the angle θ varies depending on the position occupied by Pluto in its orbit around the sun. Taking values from Table 1 we obtain a maximum θ = 1° value if Pluto has an ionopause, or θ = 0.3° if it has a magnetosphere.

3 Discussion and conclusions

The main scenarios that the New Horizons spacecraft may find at Pluto are:

*3.1 Tidal heating.* During perihelion differential tides exerted by Sun and Charon on Pluto is 5.1 % the force exerted by Charon. At aphelion it is 1.8 %. Such difference subjects the planetary body to tension and release heat that can be manifested as some kind of geological activity.

*3.2 Magnetic field.* Even though the presence of an intrinsic magnetic field in Pluto is unlikely it is still possible. If we assume that Pluto has a molten iron core (quite unlikely) then we can suggest that the magnetic moment is $5.34 \times 10^{18}$ Amp-$m^2$, an order of magnitude smaller than that of Mercury. In case that an underground saltwater ocean produces a magnetic field, as in the case of Ganymede, the magnetic moment would be $1.03 \times 10^{14}$ Amp-$m^2$. With the largest magnetic moment, the subsolar point of the magnetopause will be 5768 km from the center of Pluto, i.e., 4.95 Plutonian radius. With the smaller magnetic moment it would be by the ionopause.

*3.3 Plasmasphere and Plasmapause.* Assuming there are only protons in the Plutonian plasmasphere corotation zone has a radius of 19,100 km, for the lower magnetic moment, and 238,000 km for the higher magnetic moment. In either case the radius of corotation zone is larger than the distance to the sub-solar magnetopause, and thus the entire front of the magnetosphere co-rotates with the planet.

*3.4 Ionopause.* The height of the ionopause at the subsolar point is 2147 km and grows to 2261 km by the terminator. Measured from the center of the planet it is 3312 km, at the subsolar point, and 3426 km by the terminator.

*3.5 Trans-terminator flow.* The trans-terminator flow speed obtained from the model is 2.8 km/s, which is larger than the 1.22 km/s escape velocity. Therefore, the momentum transfer from the solar wind to the Pluto´s ionosphere should produce a significant loss of ionospheric material.

*3.6 Angle of the wake.* Pluto's plasma wake is subject to two different deviations: A) the aberration angle of arrival of the solar wind since Pluto's orbital motion is perpendicular to the radial direction of flow (at Pluto the mean angle is $\varphi = 0.6°$). B) The angle due to the Magnus force. At Pluto this angle is $\theta = 11.5°$ if Pluto has an ionopause, or $\theta = 0.3°$ if it has a magnetosphere. These are maximum values. Unfortunately, when the New Horizons spacecraft approaches Pluto, the conditions will not be optimal for observing a deviation of the Magnus angle in the wake because the solar wind direction and the rotation axis will be parallel. At Venus the aberration angle and the Magnus angle are in the same plane and thus they add up so that it is difficult to quantify the contribution of each deviation on the wake. In Pluto both angles are in nearly perpendicular planes, and thus they can be distinguished from each other.

Referen**ces**

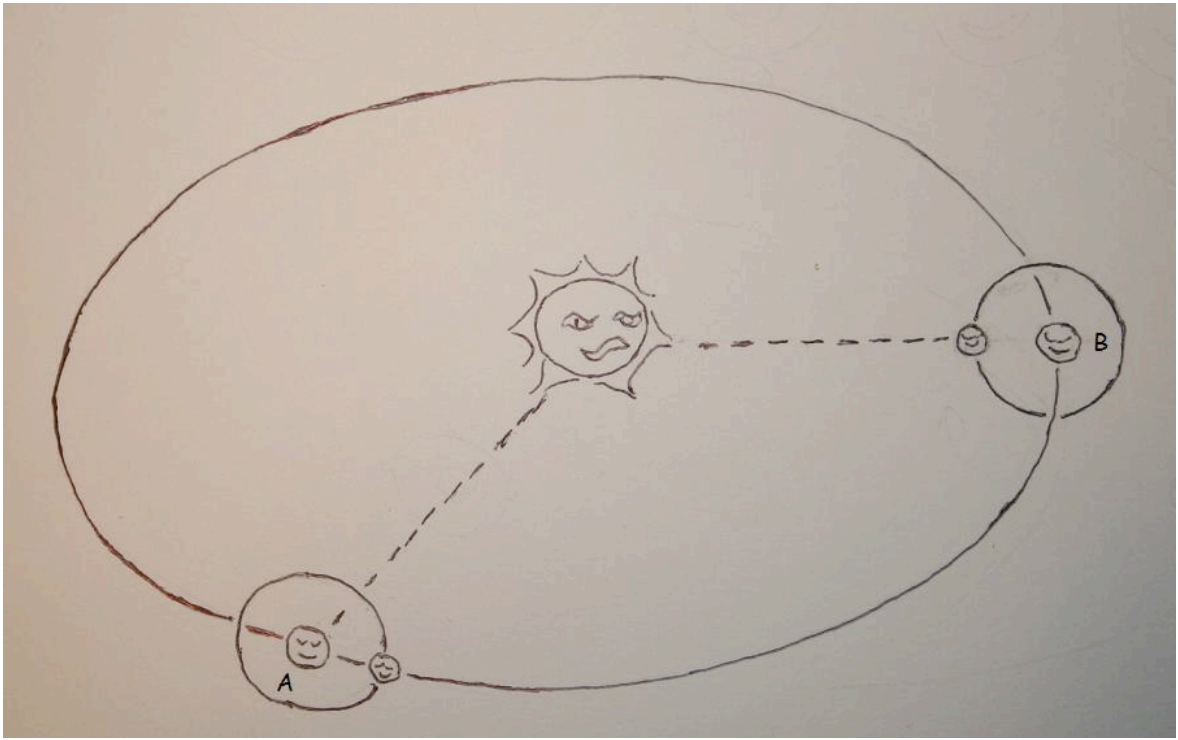

Figure 1. At point A the Sun and Charon are not aligned with Pluto, then there are no tidal forces. But at point B are aligned with Pluto at the conjunction and opposition, at that time there are tidal force.

Table 1
Values of the variables used in calculations

| Variable | Value | Reference |
|---|---|---|
| Proton charge (q) | $4.803 \times 10^{-19}$ C | Cox, 2000 |
| Gravity constant (G) | $6.672 \times 10^{-11}$ m$^3$ kg$^{-1}$ s$^{-2}$ | Cox, 2000 |
| Solar Wind density (ρ) | $1.67 \times 10^{-23}$ kg/m$^3$ | Delamere et al., 2004[1] |
| Pluto-Charon distance (Rc) | 19571.4 ± 4.0 km | Young et al., 2007 |
| Seawater conductivity (max.) | 6.5683 S/m | Kennish, 2001 |
| Conductivity of iron | $1.2 \times 10^5$ S/m | Kittel, 2004 |
| Pluto-Sun distance (Rs) (aphelion) | $7.37 \times 10^9$ km | Tholen et al., 2000[1] |
| Pluto-Sun distance (Rs) (perihelion) | $4.44 \times 10^9$ km | Tholen et al., 2000[1] |
| Charon Mass (Mc) | $1.52 \times 10^{21}$ kg | Young et al., 2007 |
| Proton Mass (m) | $1.672 \times 10^{-27}$ kg | Cox, 2000 |
| Pluto Mass (Mp) | $1.305 \pm 0.006 \times 10^{22}$ kg | Young et al., 2008 |
| Sun Mass (Ms) | $1.989 \times 10^{30}$ kg | Cox, 2000 |
| Pluto rotation period ($P_e$) | 6.38 days = $5.512 \times 10^5$ s | Buie et al., 2012 |
| Permeability of vacuum ($\mu_0$) | $1.26 \times 10^{-6}$ H/m | Resnick, 1970 |
| Pluto radius ($r_0$) | 1165 ± 25 km | Young et al., 2008 |
| Ionosphere temperature (T) | 1000 K | Ip et al., 2000 |
| Angular velocity of Pluto (ω) | $1.14 \times 10^{-5}$ rad/s | Buie et al., 2012[1] |
| Solar Wind velocity (V) | $4.5 \times 10^5$ m/s | Delamere et al., 2004 |
| Orbital velocity of Pluto ($V_o$) | 4.749 km/s | Tholen et al., 2000 |

[1] Calculation based on data from the authors cited.

Appendix
Meaning of the symbols used

$f_M$ ≡ Magnus force.
$F_C$ ≡ Gravity force of Charon on Pluto.
$F_S$ ≡ Gravity force of the Sun on Pluto.
$G$ ≡ Gravity Constant
$h$ ≡ Height above the surface of Pluto.
$k$ ≡ Boltzmann Constant.
$M$ ≡ Magnetic Moment
$M_p$ ≡ Pluto Mass
$M_C$ ≡ Charon Mass
$M_S$ ≡ Sun Mass
$m$ ≡ Plasmasphere Particles Mass
$m_p$ ≡ Proton Mass
$n_i$ ≡ Ionosphere Particle Density
$n_{sw}$ ≡ Solar Wind Particle Density
$P_e$ ≡ Planet Rotation Period
$q$ ≡ Proton Charge
$r$ ≡ Corotation region radius
$r_0$ ≡ Pluto Radius
$r_i$ ≡ Ionopause radius at terminator
$R$ ≡ Ionopause or Magnetopause Radius
$R_C$ ≡ Distance of Charon to Pluto
$R_S$ ≡ Distance of the Sun to Pluto
$R_{SS}$ ≡ Distance from the centre of Pluto to the magnetopause in the sub-solar point
$t$ ≡ time
$T$ ≡ Ionosphere Temperature
$v_p$ ≡ Rotational velocity of Pluto's equator
$v_0$ ≡ Orbital Velocity of Pluto
$V$ ≡ Solar Wind Velocity before the shock wave
$V_f$ ≡ Velocity of the flow in the Magnus force.
$V_i$ ≡ Trans-terminator flow velocity
$V_p$ ≡ Ionosphere or plasmasphere rotation velocity
$V_{SW}$ ≡ Shocked Solar Wind speed in the terminator (~ 1/8 to 1/10 of V)
$V_1$ ≡ Sum of the solar wind and rotation velocities of the ionosphere or plasmasphere
$V_2$ ≡ Difference of the solar wind and rotation velocities of the ionosphere or plasmasphere
$x_1$ ≡ Distance travelled by the solar wind, near the ionopause, on the side parallel to rotation of the ionosphere
$x_2$ ≡ Distance travelled by the solar wind, near the ionopause, on the side anti-parallel to rotation of the ionosphere

$\varphi \equiv$ Aberration angle of the solar wind
$\mu_0 \equiv$ Vacuum Permeability
$\theta \equiv$ Deflection angle of the wake due to the Magnus force
$\rho \equiv$ Solar Wind Density
$\rho_i \equiv$ Ionosphere Density
$\rho_{sw} \equiv$ Shocked Solar Wind Density (~ 8 to 10 times $\rho$)
$\sigma \equiv$ Fluid layer Conductivity
$\xi \equiv$ Subsolar angle
$\omega \equiv$ Pluto Angular velocity